\documentclass[prl,twocolumn,aps]{revtex4}
\usepackage{graphicx,soul}

\begin{document}

\title{Anderson localization and nonlinearity in one dimensional disordered
photonic lattices}

\author{Yoav Lahini$^{1}$, Assaf Avidan$^{1}$, Francesca Pozzi$^{2}$, Marc
Sorel$^{2}$, Roberto Morandotti$^{3}$, Demetrios N.
Christodoulides$^{4}$ and Yaron Silberberg$^{1}$}

\affiliation{$^{1}$Department of Physics of Complex Systems, the
Weizmann Institute of Science, Rehovot, Israel}
\email{yoav.lahini@weizmann.ac.il}

\affiliation{$^{2}$Department of Electrical and Electronic
Engineering, University of Glasgow, Glasgow, Scotland}

\affiliation{$^{3}$Institut national de la recherch$\acute{e}$
scientifique, Universit$\acute{e}$ du Qu$\acute{e}$bec, Varennes,
Qu$\acute{e}$bec, Canada}

\affiliation{$^{4}$CREOL/College of Optics, University of Central
Florida, Orlando, Florida, USA}

\date{19 April 2007; published 10 January 2008}

\begin{abstract}
We experimentally investigate the evolution of linear and
nonlinear waves in a realization of the Anderson model using
disordered one dimensional waveguide lattices. Two types of
localized eigenmodes, flat-phased and staggered, are directly
measured. Nonlinear perturbations enhances localization in one
type, and induce delocalization in the other. In a complementary
approach, we study the evolution on short time scales of
$\delta$-like wavepackets in the presence of disorder. A
transition from ballistic wavepacket expansion to exponential
(Anderson) localization is observed. We find an intermediate
regime in which the ballistic and localized components coexist
while diffusive dynamics is absent. Evidence is found for a faster
transition into localization under nonlinear conditions.


\end{abstract} \pacs{ 42.25.Dd 42.65.-k 72.15.Rn 42.65.Tg }

\maketitle The propagation of waves in periodic and disordered
structures are at the foundations of modern condensed-matter
physics. Anderson localization is a key concept, formulated to
explain the spatial confinement due to disorder of quantum
mechanical wavefunctions that would spread over the entire system
in an ideal periodic lattice \cite{And,Lee,John,Sheng}. Although
Anderson localization was studied experimentally, the underlying
phenomena - the emergence of localized eigenmodes and the
suppression of wavepacket expansion - were rarely observed
directly \cite{Micro1, Maynard}. Instead, localization was usually
studied indirectly by measurements of macroscopic quantities such
as conductance \cite{Lee}, backscattering \cite{Weak,reflected}
and transmission \cite{Weirsma, Diff}.

An interesting issue concerns the effect of nonlinearity on
Anderson localization. Nonlinear interactions between the
propagating waves and nonlinearly accumulated phases can
significantly change interference properties, thus fundamentally
affecting localization. The theoretical study of the nonlinear
problem advanced using several approaches: the study of the
transmission problem \cite{Transmission}; the study of the effect
of nonlinear perturbations on localized eigenmodes \cite{aubri};
and the study of the effect of nonlinearity on wavepacket
expansion in the presence of disorder \cite{Shepelyanski}. Only a
few experiments were reported \cite{Maynard}. Recently, optical
studies enabled the study of wave evolution in nonlinear
disordered lattices \cite{HagaiThesis,Pertch,Segev}, using a
scheme discussed in \cite{De Raedt,abdu}. In particular, Schwartz
\textit{et. al.} \cite{Segev} reported the observation of Anderson
localization of expanding wavepackets in 2D lattices.

\begin{figure}
\includegraphics[clip,width=0.85\columnwidth]{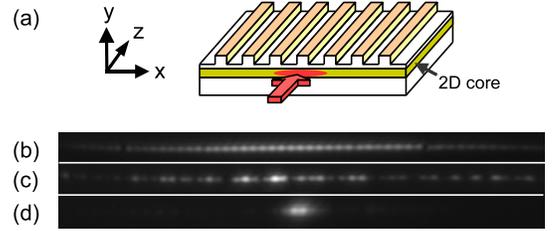}
\caption{(color online). (a) Schematic view of the sample used in
the experiments. The red arrow indicates the input beam. (b)-(d)
Images of output light distribution, when the input beam covers a
few lattice sites: (b) in a periodic lattice, (c) in a disordered
lattice, when the input beam is coupled to a location which
exhibits a high degree of expansion, and (d) in the same
disordered lattice when the beam is coupled to a location in which
localization is clearly observed.}
\end{figure}

In this work we investigate directly linear and nonlinear wave
evolution in one dimensional (1D) disordered photonic lattices,
using two different approaches. In the first part of this work,
all the localized eigenmodes of a weakly disordered lattice are
selectively excited. Nonlinearity is then introduced in a
controlled manner, to examine its effect on localized eigenmodes.
The second part of this work presents a study of the effect of
disorder on the evolution of $\delta$-like wavepackets (single
site excitations). A transition from free ballistic wavepacket
expansion to exponential localization is observed, as well as an
intermediate regime of coexistence. We then measure the effect of
nonlinearity on this process.

\begin{figure*}
\includegraphics[bb=62bp 336bp 519bp 472bp,clip,width=1.95\columnwidth]{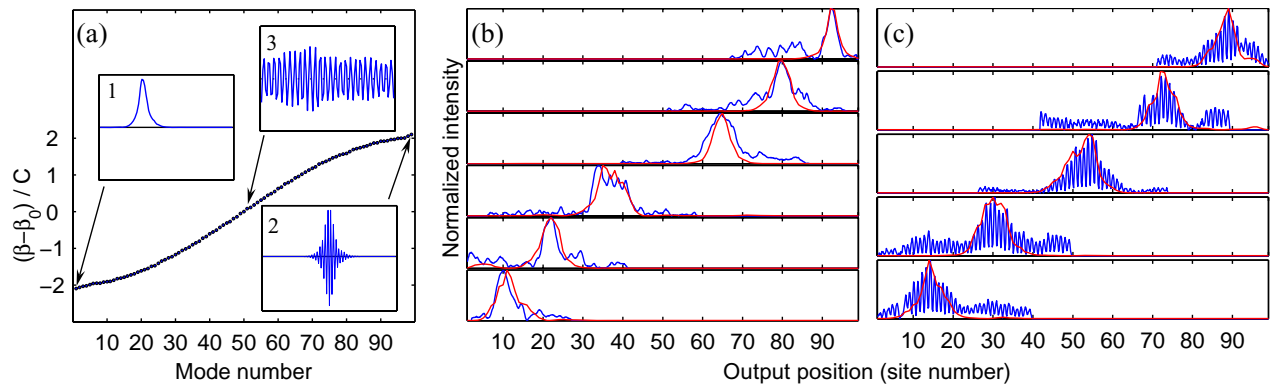}
\caption{(color online). Measurements of localized eigenmodes of a
disordered lattice. (a) Calculated eigenmodes and eigenvalues of
the weakly disordered lattice used in the experiments. The band of
eigenvalues deviates slightly from the cosine-shaped band of a
periodic lattice. Localized modes are formed, associated with
eigenvalues near the edges of the band (insets 1,2), while modes
near the band center remain extended (inset 3). (b) Measurements
of pure flat-phase Anderson localized modes. Panels show a
comparison between measurements (blue) and the corresponding
calculated eigenmodes of the lattice (red). (c) Same for staggered
localized eigenmodes. In all cases no fitting procedures are
used.}
\end{figure*}

Our experimental setup is a one-dimensional lattice of coupled
optical waveguides patterned on an AlGaAs substrate
\cite{DSoliton,Christ}, illustrated in Fig. 1a. Light is injected
into one or a few waveguides at the input and can coherently
tunnel between neighboring waveguides as it propagates along the z
axis. Light distribution is then measured at the output (see for
example Fig. 1(b)-(d)).

The equations describing light dynamics in these structures are
identical (in the linear limit) to the equations describing the
time evolution of a single electron in a lattice under the tight
binding approximation \cite{Christ}, i.e. a set of coupled
discrete Schrodinger equations:
\begin{equation}
-i\frac{\partial{U_{n}}}{\partial{z}}=\beta_{n}U_{n}+C_{n,n\pm1}\left(U_{n+1}+U_{n-1}\right)+\gamma|U_{n}|^{2}U_{n}
\end{equation}.
Here $n=1,...,N$ where $N$ is the number of lattice sites
(waveguides), $U_{n}$ is the wave amplitude at site $n$,
$\beta_{n}$ is the eigenvalue (propagation constant) associated
with the n'th site , $C_{n,n\pm1}$ are the tunnelling rates
between two adjacent sites, and $z$ is the longitudinal space
coordinate. The last term in Eq. (1) describes the nonlinear
dependence of the refractive index on the light intensity, where
$\gamma$ is the Kerr nonlinear parameter, which is positive for
our system $(\gamma>0).$ The nonlinear term can be discarded for
low light intensities. for typical experimental parameters see for
example \cite{DSoliton}.

Disorder is introduced to the lattice by randomly changing the
width of each waveguide in a finite range $W\pm\delta$ where $W$
is the mean value (typically 4$\mu m$ in our samples). The
parameters $\beta_{n}$ and $C_{n,n\pm1}$ can be calculated
numerically from the waveguides' width and from the separation
between neighboring waveguides. As a result of disorder the
parameters $\beta_{n}$ become random in the range
$\beta_{0}\pm\Delta$. By keeping the lattice periodic on average
(the site's centers still have the lattice periodicity), the
parameters $C_{n,n\pm1}$ become independent of $n$ to a very good
approximation, meeting the conditions assumed by Anderson in his
original model (diagonal disorder) \cite{And}. A measure of
disorder is then given by the ratio $\Delta/C$ \cite{Sheng}.
\begin{figure}[b]
\includegraphics[bb=190bp 355bp 426bp 447bp,width=0.9\columnwidth]{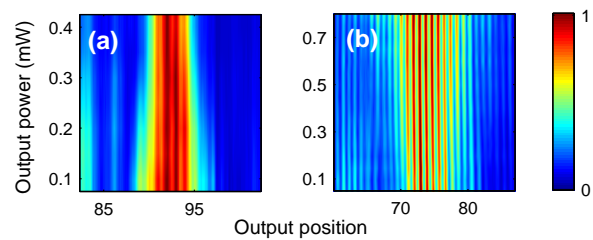}
\caption{(color online). The effect of weak nonlinearity on pure
localized eigenmodes: cross-sections of the output light
intensities (horizontal axis) at different power levels (vertical
axis), showing that (a) flat phased localized modes tend to become
more localized as nonlinearity is increased, while (b) staggered
localized modes tend to delocalize. All cross-sections are
normalized to unit maximum. }
\end{figure}
\begin{figure*}
\includegraphics[bb=67bp 310bp 537bp 480bp,width=1.9\columnwidth]{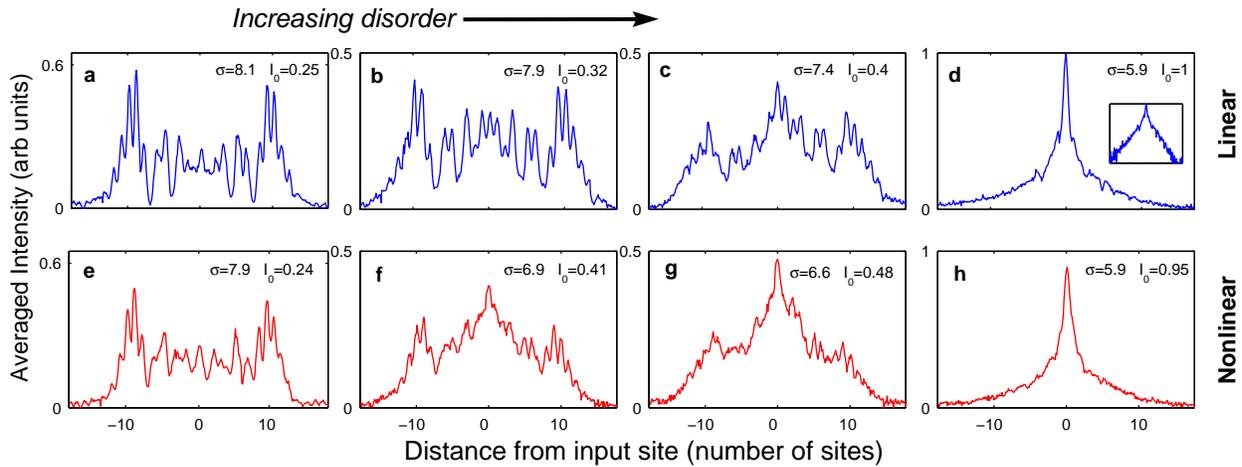}

\caption{(color online). The effect of disorder on wavepacket
expansion in the linear and the nonlinear regimes. The results
shown are normalized lattice averages of the output light
distribution, when initially a single site is excited (see
discussion in the text). A measure of localization is given by the
mean square displacement from the input site $\sigma$, and by the
intensity at the origin site $I_{o}$. (a-d) Measurements in the
linear case show the transition from ballistic transport to
exponential (Anderson) localization as a function of disorder in:
(a) $\Delta/C=0$, (b) $\Delta/C=1$, (c) $\Delta/C=1.5$, and (d)
$\Delta/C=3$. Note the transition from ballistic transport in (a)
to localization in (d) through the buildup of a central component
and the suppression of the ballistic side lobes. Inset in (d)
shows the localized distribution in semilog scale, demonstrating
the exponential tails. (e)-(h): Measurements of the same lattices
in the nonlinear case, showing that on average nonlinearity tends
to increase localization for intermediate disorder levels (e.g.
compare (b) and (f)).}
\end{figure*}

We now turn to the first set of experiments, designed to excite
and measure individually all the localized eigenmodes of a
disordered lattice, and to study the effect of nonlinear
perturbations on them. For this purpose we have fabricated a
lattice with $N$=99 sites and disorder level $\Delta/C$=1. The
disordered lattice eigenmodes and eigenvalues can be calculated by
diagonalizing a set of N equations (1) in the linear limit,
substituting the calculated values of $\beta_{n}$ and
$C_{n,n\pm1}$ for this specific realization. Results are shown in
Fig 2(a). The resulting band of eigenvalue deviates only slightly
from the cosine-shaped band of a perfectly ordered lattice
\cite{Christ}. Several eigenmodes with eigenvalues near the edges
of the band become exponentially localized in space, even though
the density of states near the band edges is not significantly
lower from the one at the band center \cite{Sheng}. Localized
eigenmodes near the bottom of the band are flat phased, i.e. their
wavefunction's amplitude is in-phase at all sites (see inset
1)\cite{Rem}, while the localized eigenmodes at the top of the
band are staggered, i.e. their wavefunction's amplitude has a
$\pi$ phase flip between adjacent sites (inset 2). Notably, These
localized eigenmodes are localized at well-separated regions in
space, and have a similar width of about 10 lattice sites.
Eigenmodes near the band center remain extended in the finite
sized system (see inset 3). These modes will also be localized in
an infinite system, but on a much larger length scale
\cite{POnAverage}. As disorder ($\Delta/C)$ is increased, a larger
fraction of the modes becomes localized within the finite lattice.


In order to excite localized eigenmodes of this lattice we inject
a wide beam (covering several lattice sites) at different
locations across the lattice. At some input positions, we observe
significant expansion of the beam at the output, similar in width
to the expansion observed in a periodic lattice with the same
average parameters (see fig 1(b) and 1(c)). At specific locations
however, wave expansion is suppressed and prominent localization
is evident (Fig 1(d)). In these cases, the input beam overlap
significantly with a single localized eigenmode of the lattice,
while the overlap with the other modes of the lattice is
eliminated. Optimized localized output distribution were achieved
using an input beam that covers about 10 lattice sites,
corresponding to the typical width of the localized eigenmodes in
this disorder level.

While scanning the input position we observe several localized
light distributions at the output. We compare the obtained
localized intensity profiles to calculated localized eigenmodes of
the lattice in Fig 2(b). There is a clear correspondence between
the experimentally observed localizations and the location and
shape of all the calculated flat-phased localized eigenmodes
associated with the bottom of the eigenvalue band. To excite the
staggered modes associated with the top of the band, the input
beam was tilted with respect to the lattice to induce a $\pi$
phase difference in the excitation of adjacent waveguides
\cite{Christ}. The results of this excitation scheme are presented
in Fig. 2(c). Again, a clear correspondence is found to the
calculated localized staggered modes of the lattice. These results
demonstrate the ability to excite pure Anderson localized
eigenmodes.



The effect of nonlinear perturbations on localized eigenmodes is
studied by exciting a pure localized mode and increasing the input
beam power. The intensities are kept way below those required to
exhibit self-focusing in a periodic lattice with the same average
parameters \cite{DSoliton}, keeping the experiments in the weak
nonlinear regime. Some localized modes are found to exhibit
significant response to nonlinearity. The results of two such
experiments are shown in Fig. 3, showing that weak positive
nonlinearity tends to further localize flat-phased localized
modes, but tends to de-localize staggered modes.

These results can be understood on the basis of the theory
developed in \cite{aubri}, which have shown that nonlinear shifts
of a localized eigenmode's frequency (represented here by the
parameter $\beta$), can lead to delocalization if the frequency
crosses a resonance with other modes of the lattice. This
condition can be satisfied in the case of weak disorder for the
staggered localized modes, as nonlinearity shifts them from the
edge of the band into the linear spectrum. Conversely, the
flat-phased modes at the other edge of the band are shifted by
nonlinearity out of the linear spectrum, thus they remain
localized (see also discussion in \cite{Pertch}).

We now turn to the second set of experiments, aimed to study the
effect of disorder on wavepacket evolution in the linear and
nonlinear regimes. This aspect is studied by injecting light into
a \textit{single} lattice site, thus exciting a tight
$\delta$-like wavepacket of all eigenmodes having non-vanishing
overlap with the excited site. The wavepacket then evolves in the
lattice, and the light distribution is measured at the output. We
average the output patterns obtained, by separately exciting each
site in the lattice while keeping the measurement-window centered
about the input site position. The results of such measurements in
the linear regime, taken in 5mm long samples with different
disorder levels, are shown in Fig. 4(a-d). Without disorder,
single site excitation results in ballistic propagation
(wavepacket width grows linearly with time), recognized by a
characteristic signature of two separated lobes
\cite{Christ,DSoliton} (Fig. 4(a)). At moderate disorder, a second
component emerges, localized around the input site position (Fig.
4(b,c)). The localized and the ballistic components coexist in
this regime. At high disorder a highly localized, exponentially
decaying distribution is observed (Fig. 4(d)). This exponential
decay of the expansion profile is a hallmark of Anderson
localization.

These results offer a first direct look at the short time
evolution of wavepackets in 1D disordered systems. It is known
that for infinite disordered 1D systems and for long time scales,
wavepacket expansion is always fully suppressed. However, on short
time scales, wavepackets do evolve \cite{De Raedt,Evolution}. The
results in Fig. 4(a-d) reveal how localization is reached in the
1D case; localization emerges from ballistic expansion through the
continues buildup of a localized component and the suppression of
a ballistic component. This dynamics is fundamentally different
from the one observed in 2D systems \cite{Pertch, Segev}, or that
was indirectly measured in quasi-1D experiments \cite{Diff}. In
these cases the expansion turns quickly from ballistic to
diffusive, and becomes localized after much longer propagation
times. In the 1D case the diffusive dynamics is absent, as
discussed for example in \cite{Evolution}.

To study the effect of nonlinearity on wavepacket expansion we
repeat these measurements at increased light intensities. Again,
we remain in the weak nonlinear regime. Results are shown in Fig.
4(e-h). On average, the results indicate increased localization at
intermediate disorder levels; the intensity in the ballistic
component is lower and the intensity in the localized component is
higher under nonlinear conditions. This suggests that the buildup
of the localized component and the suppression of the ballistic
component happen faster under nonlinear conditions. This
description holds for short time scales in which the ballistic
component is still present. On much longer time scales,
subdiffusive delocalization due to nonlinearity was predicted to
take over \cite{Shepelyanski}.

%

In conclusion, we have directly studied localized eigenmodes and
wavepacket expansion in disordered 1D lattices that are described
by the nonlinear version of the Anderson model. Two types of
localized eigenmodes were measured, and nonlinear perturbations
were shown to enhances localization in one type and induce
delocalization in the other. The study of the expansion of
wavepackets on short time scales in the presence of disorder has
enabled a direct measurement of the transition from ballistic
wavepacket expansion to exponential localization. It was shown
that in 1D systems a ballistic and a localized components can
co-exist at intermediate times, while diffusive wavepacket
expansion, observed in 2D and quasi-1D systems, is absent in the
1D case. In the nonlinear regime, evidence is found for a faster
transition into localization under nonlinear conditions.

We thank Y. Imry, M. Aizenman and H.S. Eisenberg for useful
discussions. This work was supported by the German-Israeli Project
Cooperation (DIP), NSERC and CIPI (Canada), and EPRSC (UK). Y.L.
is supported by the Adams Fellowship Program of the Israel Academy
of Sciences and Humanities.


\begin{thebibliography}{10}
\bibitem{Rem}The band here is inverted in respect to the usual convention
in waveguide lattices.
\end{thebibliography}
\end{document}